\documentclass[12pt]{article}

\usepackage[totalheight = 23cm, totalwidth = 17cm]{geometry}
\usepackage{amssymb,amsmath,amsfonts,amsbsy,graphicx,bm}

\usepackage{color,cancel,ulem}

\newcommand{\dis}[1]{\begin{equation}\begin{split}#1\end{split}\end{equation}}

\begin{document}

\begin{titlepage}

\begin{center}

{\LARGE \bf 
Dilaton stabilization in KKLT revisited
}

\vskip 1.0cm

{\large
Min-Seok Seo$^{a}$ 
}

\vskip 0.5cm

{\it
$^{a}$Department of Physics Education, Korea National University of Education,
\\ 
Cheongju 28173, Republic of Korea
}

\vskip 1.2cm

\end{center}

\begin{abstract}

 We study the condition for the dilaton stabilization in Type IIB flux compactifications consistent with the KKLT scenario.
 Since the Gukov-Vafa-Witten superpotential depends linearly on the dilaton, the dilaton mass squared is given by a sum of the gravitino mass squared and additional terms determined by the complex structure moduli stabilization.
 If the dilaton mass is not much enhanced from the gravitino mass, the mass mixing with the K\"ahler modulus in the presence of the non-perturbative effect generates the saddle point at the supersymmetric field values, hence the potential becomes unstable.
 When the complex structure moduli other than the conifold modulus are neglected, the saddle point problem arises over the controllable parameter space.
 We also point out that the dilaton stabilization condition is equivalent to the condition on the NS 3-form fluxes, $|H_{(1,2)}| > | H_{(0,3)}|$.

\end{abstract}

\end{titlepage}

\newpage

\section{Introduction}

 The construction of convincing cosmological models from string theory has been a challenging issue for more than two decades.
 This essentially requires compactifications to four dimensions in which all the moduli are fixed at the metastable de Sitter (dS) vacuum. 
 The stabilization of the dilaton and the complex structure moduli was shown to be possible by turning on 3-form fluxes in Type-IIB Calabi-Yau orientifold compactifications with O3/O7-planes, which we will call the Giddings-Kachru-Polchinski (GKP) solution \cite{Giddings:2001yu} (see also \cite{Dasgupta:1999ss} for an example of earlier flux compactification).  
 On the other hand, the K\"ahler modulus associated with the  internal compactification volume remains unfixed at tree level, as reflected in the no-scale structure.  
  The Kachru-Kallosh-Linde-Trivedi (KKLT) scenario \cite{Kachru:2003aw} suggested that when the tree level superpotential is  comparable in size to the exponentially small non-perturbative quantum effect, the K\"ahler modulus can be stabilized at the supersymmetric anti-de Sitter (AdS) minimum.
 Supersymmetry is broken by putting anti-D3 ($\overline{\rm D3}$)-branes at the tip of the Klebanov-Strassler (KS)-like throat \cite{Klebanov:2000hb}, provided the number of $\overline{\rm D3}$-branes is tuned to be much smaller than the flux quanta \cite{Kachru:2002gs}.
 Then the minimum of the potential is uplifted to a  metastable dS vacuum.
 
 The description above implies that the KKLT scenario does not work without fine-tuning of parameters, which indeed seems to be a common feature of string models for the metastable dS vacuum including the Large Volume Scenario \cite{Balasubramanian:2005zx}. 
 Presumably, there may be some as yet unknown reason in string theory (or in quantum gravity) which forbids the metastable dS vacuum  over a wide range of (or even whole) parameter space as argued in the ``dS swampland conjecture" \cite{Obied:2018sgi} (see \cite{Garg:2018reu, Andriot:2018mav, Ooguri:2018wrx} for a refined version, especially \cite{Ooguri:2018wrx} and also \cite{Seo:2019mfk, Seo:2019wsh} for attempts to find out the thermodynamic argument to support the conjecture).
 Such a concern has motivated consistency checks for the KKLT scenario in various directions, investigating whether the scenario works  over the controllable parameter space of the effective supergravity (e.g., \cite{McGuirk:2009xx, Bena:2009xk, Michel:2014lva, Moritz:2017xto, Bena:2018fqc, Hamada:2018qef, Carta:2019rhx, Blumenhagen:2019qcg, Kachru:2019dvo, Randall:2019ent, Dudas:2019pls, Gao:2020xqh, Bena:2020xrh} and especially  \cite{Gao:2020xqh} for a summary on main issues and relevant references).

 What makes the problem more complicated is that  while the KKLT scenario is a combination of several mechanisms, each of them is not necessarily completely decoupled from others \cite{deAlwis:2005tf}.
The dilaton or some of the complex structure moduli  can be so light that they are not decoupled from the stabilization mechanism of the K\"ahler modulus.
For example, in \cite{Blumenhagen:2019qcg}, the strongly warped regime was explored, in which not only the K\"ahler modulus but also the conifold modulus, the complex structure modulus associated with the size of the 3-cycle of the deformed conifold \cite{Candelas:1989js} have exponentially suppressed masses.
On the other hand, the  Gukov-Vafa-Witten (GVW) superpotential \cite{Gukov:1999ya} induced by the 3-form fluxes depends linearly on  the dilaton. 
As a result, the dilaton mass squared is written as a sum of the gravitino mass squared and other terms determined by the stabilization of other moduli.
In other words, whether the dilaton is heavier than the gravitino is closely connected to how  the complex structure moduli  are stabilized by  fluxes.
 If the dilaton mass is not considerably larger than the gravitino mass, the mass mixing between the K\"ahler modulus and the dilaton needs to be carefully addressed since  the K\"ahler modulus mass is  enhanced from the gravitino mass.
 As can be inferred from \cite{Choi:2004sx}, the mass mixing in this case may generate an unstable direction such that  a set of field values obtained from the supersymmetric condition is in fact on the saddle point.
   The purpose of this article is to explore this possibility in detail.
   Since the dilaton linearly couples to any complex structure moduli in the GVW superpotential,  we expect that our investigations provide more stringent constraints on the stabilization of the complex structure moduli with fluxes which are not evident when we just focus on the conifold modulus.

 This article is organized as follows. 
 In section \ref{sec:Review}, we review main features of the moduli stabilization in the original KKLT and   GKP  which assumed that the K\"ahler modulus is decoupled from the dilaton and the complex structure moduli.
 From this, we summarize the ingredients needed for our discussion and settle the notations.
 In section \ref{sec:EFT}, we illustrate the saddle point problem by considering the extreme case in which the conifold modulus is the only complex structure modulus or other complex structure moduli effects are negligibly small, as  frequently employed in the literatures for simplicity.
  For this purpose we examine the low energy effective superpotential induced by integrating out the conifold modulus.
  Then we apply the results of \cite{Choi:2004sx} to show that the mass mixing between the dilaton and the K\"ahler modulus in the presence of the non-perturbative effect easily destabilizes the potential.
  This strongly suggests that the dilaton stabilization requires effects from complex structure moduli other than the conifold modulus.
 In section \ref{sec:Condition} we interpret the condition for the absence of the saddle point in light of the GVW superpotential in two equivalent  ways : 1) the dilaton must be considerably heavier than the gravitino mass and 2) the $(1,2)$ component of the NS 3-form fluxes dominates over the $(3,0)$ component which is consistent with the low energy supersymmetry breaking.
  This provides a more stringent constraint on the flux vacua which also reflects the effects from the large number of complex structure moduli.
Finally we conclude.
Throughout this work,  we take the (reduced) Planck unit $m_{\rm Pl}=1$ unless otherwise stated.
Notations and conventions  follow \cite{Blumenhagen:2013fgp}.

\section{Moduli stabilization in KKLT and GKP}
\label{sec:Review}

In this section we first briefly review the K\"ahler modulus stabilization in the original KKLT scenario which assumed that the dilaton and the complex structure moduli are already fixed through the GKP solution. 
We also write down the K\"ahler potential and the GVW superpotential used in the GKP solution, focusing on the dilaton and the conifold modulus.
In both cases, the moduli and the dilaton are stabilized through the F-term potential  
\dis{V=e^{K}\Big(G^{I\overline{J}}D_I W D_{\overline J}\overline{W}-3|W|^2 \Big),}
 where $G^{I\overline{J}}$ is the inverse of the K\"ahler metric $G_{I\overline{J}}=\partial_I \partial_{\overline J}K$ and $D_I W=\partial_I W+ \partial_I K W$.

\subsection{KKLT scenario}

 In the KKLT scenario \cite{Kachru:2003aw}, the K\"ahler potential and the superpotential for the K\"ahler modulus $\rho$ are given by
 \dis{K_\rho=-3\log[-i(\rho-\overline{\rho})],\quad\quad
 W_\rho=W_0+C e^{i a \rho},\label{eq:rhodyn}}
 before the uplift.
If $W_0$ and $C$ have the same phase and $a$ is real, the F-term potential has a supersymmetric AdS minimum with $\rho=i \sigma$ (i.e., a scalar and a pseudoscalar parts are fixed at $\sigma$ and $0$, respectively), satisfying $D_\rho W=0$ or
 \dis{W_0=-e^{-a \sigma} C\Big(1+\frac23 a \sigma\Big).}
 For higher-order non-perturbative effects to be negligible, $a\sigma > 1$ is required, which means that the scenario works provided $W_0$ is exponentially suppressed.
 The mass for $\sigma$ is given by
  \dis{m_\sigma^2=G^{\rho \overline{\rho}}\partial_\sigma^2 V=\frac{2 a^3 W_0^2 }{a \sigma}\Big(\frac{2+ 5(a\sigma)+2 (a \sigma)^2}{(3+2 a \sigma)^2}\Big).}
 The pseudoscalar part of $\rho$  has a mass $m_{\rm Re \rho}^2 =2 a^3 W_0^2/(3+2 (a \sigma))$, which is similar in size to $m_\sigma^2$ for $ a\sigma \gg 1$.
  After the uplift, the supersymmetry is broken to give the gravitino mass
  \dis{m_{3/2}^2=e^K |W|^2=\frac{a^2 C^2}{18\sigma}e^{-2 a\sigma},}
  which is roughly $m_{3/2}^2 \sim W_0^2/\sigma^3$ while  $m_\sigma^2$ is not much affected, hence 
  \dis{m_\sigma \sim (a \sigma) m_{3/2} > m_{3/2}.}

  \subsection{GKP solution}

 The premise of the KKLT scenario is that the dilaton $\tau=C_0+i e^{-\phi}$ as well as the complex structure moduli $t^a$ are stabilized by the flux-induced superpotential as suggested in GKP \cite{Giddings:2001yu}.
 Here we must require that at the stabilized field values,  the string coupling constant is perturbatively small $g_s=e^{\langle \phi \rangle} < 1$ and $W_0$ is  tuned to be  exponentially suppressed.
  In the original KKLT scenario, it was assumed that $\tau$ and $t^a$ are heavier than $\rho$ such that their stabilization is not much affected by the non-perturbative effect.
 However, this is not always guaranteed and as we will see, if   $\tau$ is as light as $m_{3/2}$   the vacuum is typically destabilized by the mass mixing between $\rho$ and $\tau$.

  To see this, we first review the structure of the K\"ahler potential and the superpotential in the GKP solution.
 The K\"ahler potential for $(\tau, \rho, t^a)$  is written as
 \dis{K_\tau +K_\rho +K_{\rm cs}=-\log[-i(\tau-\overline{\tau})]-3\log[-i(\rho-\overline{\rho})]-\log\Big(i \int_X \Omega\wedge \overline{\Omega}\Big),\label{eq:Kahlertau}}
 whereas the flux-induced superpotential is given by the GVW superpotential,
 \dis{W_{\rm GVW}= c  \ell_s^{-2} \int_X \Omega \wedge G_3.\label{eq:GVW}}
Here $\Omega$ is the holomorphic 3-form and $G_3= F_3-\tau H_3$.
 Since  $W_{\rm GVW}$ depends on $\tau$ through $G_3$, it is linear in $\tau$.
 A coefficient $c$ is determined by  matching  the F-term potential induced by $W_{\rm GVW}$  with the dimensional reduction of the Type-IIB supergravity action for $G_3$, which turns out to be $c=g_s^{3/2}(4\pi)^{-1/2}$ \cite{Conlon:2005ki}. 
 The value of $W_{\rm GVW}$ at the stabilized field values will be identified with $W_0$ in the KKLT scenario.
 %Here and thereafter, we take the (reduced) Planck unit $m_{\rm Pl}=1$ unless otherwise stated.

 In order to see the explicit form of $K_{\rm cs}$ and $W_{\rm GVW}$, we introduce a basis of the de Rham cohomology group $H^3 (X, \mathbb{Z})$ given by generators  $\alpha_I$ and $\beta^I$ ($I=0, \cdots h^{2,1}(X)$), which are Poincar\'e dual to the homology basis $(A^I, B_I)$ of $H_3(X, \mathbb{Z})$.
 They satisfy
 \dis{& A^I \cdot A^J=B_I \cdot B_J=0,\quad A^I \cdot B_J=\delta^I_J,
 \\
 &\int_{A^J}\alpha_I=-\int_{B_I}\beta^J=\int_X \alpha_I \wedge \beta^J=\delta_I^J.}
 In terms  of $A$- and $B$-periods of $\Omega$
 \dis{Z^I=\int_{A^I}\Omega,\quad\quad {\cal F}_I=\int_{B_I}\Omega,}
$\Omega$ is written as
\dis{\Omega=Z^I \alpha_I-{\cal F}_I \beta^I.\label{eq:Omega}}
 Then ${\cal F}_I$ corresponds to $\partial_I {\cal F}$, in which ${\cal F}(Z)$ is connected to the prepotential $F(t^a)$ through ${\cal F}(Z)=(Z^0)^2F(t^a)$ with $t^a=Z^a/Z^0$ ($a=1, \cdots h^{2,1}(X)$).
 From this, the K\"ahler potential for the complex structure moduli is written as
 \dis{K_{\rm cs}=-\log\Big(i \int_X \Omega\wedge \overline{\Omega}\Big)=-\log\Big(i|Z^0|^2\Big[2(F-\overline{F})-(t^a-\overline{t}^a)(F_a+\overline{F}_a)\Big]\Big).\label{eq:Kahler0}}
Meanwhile,  fluxes $F_3$ and $H_3$ are quantized with respect to $H^3 (X, \mathbb{Z})$ such that
\dis{F_3=\ell_s^2 (M^I \alpha_I+M_I \beta^I)\quad\quad 
H_3=\ell_s^2 (K^I \alpha_I+K_I \beta^I).}
The quanta $(M^I, M_I)$ and $(K^I, K_I)$ are restricted by the tadpole cancellation condition,
\dis{\frac{1}{\ell_s^4}\int_X F_3 \wedge H_3=M^IK_I-M_I K^I =-N_{D_3}+\frac12 N_{O_3}.\label{eq:tadpole}}
In the presence of the wrapped D7-brane, the D3 charge $\chi(X)/24$ can be induced, where $\chi(X)$ is the Euler characteristic of the Calabi-Yau 4-fold in the F-theory compactifications \cite{Sethi:1996es}.
This value is added to the r.h.s. of \eqref{eq:tadpole}.

In the GKP solution, one of the complex structure moduli corresponds to the conifold modulus $z$, a period of a three-cycle $A$ which vanishes as  $S^3$ of a cone $S^3 \times S^2$ shrinks to zero size.
\footnote{More precisely, the deformed conifold is described by four complex coordinates $w_i$ ($i=1, \cdots 4$) satisfying $\sum_i w_i^2=\epsilon^2$.
The conifold modulus $z$ determines the deformation parameter $\epsilon$ through the relation $\epsilon^{8/3}={\alpha'}^2[-i(\rho-\overline{\rho})]|z|^{4/3}$.
The $\rho$ dependence here will be discussed later. \label{footnote:con}}
Setting $Z^0=1$, one finds that
\dis{\int_A \Omega =z \quad\quad \int_B \Omega ={\cal F}_z=\frac{z}{2\pi i}\log z +({\rm holomorphic}).\label{eq:zcycle}}
Let us assign the flux charges relevant to $(A, B)$ as
\dis{\int_A F_3 = \ell_s^2 M,\quad\quad \int_B H_3 =-\ell_s^2 K,}
i.e., $(M^z, M_z)=(M, 0)$ and $(K^z, K_z)=(0, K)$.
Moreover, we assume that the prepotential has the form of $F(z, t^i)=F_1(z)+F_2(t^i)$ with $i=2, \cdots h^{2,1}(X)$, such that
\dis{F_1(z)=\frac{z^2}{4\pi i}\Big(\log z-\frac12\Big)+f_0 +f_1 z+\cdots.}
Then the K\"ahler potential is written as
\dis{K_{\rm cs}=-\log\Big(&\frac{|z|^2}{2\pi}\log |z|^2+2i(f_0-f_0^*)+i(f_1-f_1^*)(z+{\overline z})+\cdots
\\
& +i[2(F_2-\overline{F}_2)-(t^i-\overline{t}^i)(F_{2,i}+\overline{F}_{2,i})]\Big),\label{eq:Kahler1}}
where $\cdots$ includes higher order terms (${\cal O}(z^2)$) that vanish as $z\to 0$.
Meanwhile, since ${\cal F}_0=Z^0(2F-z F_z-t^i F_i)$,  the superpotential with $Z^0=1$ is written in the form of
\dis{W_{\rm GVW}=A(z, t^i)-\tau B(z, t^i),}
where
\dis{A(z, t^i)&=c(Z^I M_I+F_I M^I) 
\\
&=c\Big( M_0+(2 f_0+ f_1 z)M^0+\big(\frac{z}{2\pi i}\log z+f_1\big)M+(2 F_2-t^i F_i)M^0+ t^iM_i+F_i M^i\Big),
\\
B(z, t^i)&=c(Z^I K_I + F_I K^I)
\\
&=c\Big(K_0+ (2 f_0+ f_1 z)K^0+ z K + (2 F_2-t^i F_i)K^0+t^i K_i+F_i K^i\Big).}
Here terms higher order in $z$  are omitted as well.
We note that even if $t^i$ are integrated out, the effective superpotential should not contain $z$ dependent terms  enhanced from $z\log z$ and $\tau z$  since otherwise $z$ cannot be stabilized at the exponentially small value.
The separation of $z$ and $t^i$ in both the K\"ahler potential and the superpotential is helpful for this purpose. 
While the discussion so far is generic, a concrete example, e.g., the mirror dual of the quintic $\mathbb{P}_4[5]^{(101,1)}$ gives the K\"ahler potential and the superpotential of the same structures as those we have shown \cite{Blumenhagen:2016bfp} (see also  \cite{Candelas:1990rm, Curio:2000sc, Huang:2006hq}).

  If $\tau$ as well as $t^a=(z, t^i)$ are heavy enough to be decoupled from the KKLT scenario, the non-perturbative effect can be ignored in their stabilization.
 In this limit, since the K\"ahler potential for $\rho$ is given by $K_\rho$ in \eqref{eq:Kahlertau} and $W_{\rm GVW}$ is independent of $\rho$, the F-term potential has a no-scale structure, $V=e^K G^{I\overline{J}}D_I W D_{\overline{J}}\overline{W}$, where the indices $K,~L$ run over fields other than $\rho$. 
 Then the second derivative matrix of the potential at the  minimum satisfying $D_I W=0$ consists of
 \dis{\partial_I\partial_J V=e^K G^{K\overline{L}}[\partial_I D_K W \partial_J D_{\overline L} \overline{W} + \partial_J D_K W \partial_I D_{\overline L} \overline{W}].\label{eq:v''}} 

 Meanwhile,  the fluxes and  the branes backreact on the geometry, resulting in the warped geometry in which the metric around the throat is given by
   \dis{ds^2=e^{2A_0}\eta_{\mu\nu}dx^\mu dx^\nu+e^{-2 A_0}\widetilde{g}_{mn} dy^m dy^n.}
% When the D3 (or $\overline{\rm D3}$)-branes are located at the tip of the KS throat, the warp factor  in terms of the distance from the tip $r$ is given by \cite{Klebanov:2000nc}
% \dis{e^{-4A_0}=1+\Big(4\pi {\alpha'}^2 g_s N+c_2 (\alpha'g_s M)^2\Big[1+4\log\Big(\frac{r}{r_{\rm UV}}\Big)\Big]\Big) \frac{1}{r^4},\label{eq:warp} }
% where the Regge slope $\alpha'$ is related to the string length $\ell_s$ by $\ell_s=2\pi\sqrt{\alpha'}$, $N$ is the total D-brane charge stored in the fluxes and branes ($MK+N_{\rm D_3}$), and $c_2$ is an ${\cal O}(1)$ coefficient.
  In the dilute flux limit ($e^{-4A_0} \simeq 1$)  the internal geometry can be treated as if the Calabi-Yau manifold.
 In contrast, when the throat is highly warped ($e^{-4A_0}\gg 1$), the warp factor near the throat is given by \cite{Klebanov:2000hb}
 \dis{&e^{-4A_0}\simeq 2^{2/3} \frac{(\alpha' g_s M)^2}{\epsilon^{8/3}}{\cal I}(y),
 \\
 &{\cal I}(y)=\int_y^\infty dx \frac{x \coth(x)-1}{\sinh^2 x}(\sinh(2x)-2x)^{1/3},}
 where  $\alpha'=\ell_s^2/(2\pi)^2$ and $y$ is a coordinate along the throat : $y=0$ corresponds to the tip of the throat and ${\cal I}(y=0)$ converges to a constant.
 We infer from the metric of the KS throat \cite{Klebanov:2000hb} that the deformation parameter $\epsilon^2$  (see footnote \ref{footnote:con}) is proportional to $|z|$.
 Moreover, from the equation of motion
 \dis{\tilde{\nabla}^2 e^{4A}=e^{2A}\frac{G_{mnp}\overline{G}^{mnp}}{12{\rm Im}\tau}+\cdots,}
 where $\tilde{\nabla}^2$ is the Laplacian with respect to the unwarped internal (Calabi-Yau) metric $\widetilde{g}_{mn}$, one finds that  under $g_{mn} \to \lambda g_{mn}$, the warp factor for the throat part also scales as $e^{-4A_0} \to  \lambda^{-2} e^{-4A_0}$ while the overall volume $V_X =\ell_s^6[-i(\rho-\overline{\rho})]^{3/2}$ scales as $V_X \to \lambda^3V_X$.
 \footnote{The scaling behavior of $V_X$ comes from
  \dis{V_X=\int d^6y \sqrt{\tilde{g}}e^{-4A_{\rm tot}}.}
 Here $e^{-4A_{\rm tot}}$ is  the warp factor for the {\it overall internal volume}, which should not be confused with the warp factor for the {\it throat} $e^{-4A_0}$.
}
This suggests   the relation $\epsilon^{8/3}={\alpha'}^2[-i(\rho-\overline{\rho})] |z|^{4/3}$  \cite{Giddings:2005ff}. 
 In this strongly warped regime  $W_{\rm GVW}$ is not changed but $K_{\rm cs}$ in \eqref{eq:Kahler0} is modified to be \cite{DeWolfe:2002nn}
\dis{K_{\rm cs}=-\log\Big(i\int_X e^{-4A_0} \Omega\wedge \overline{\Omega}\Big),}
such that the K\"ahler metric for $z$ has the form of \cite{Douglas:2007tu}
\dis{G_{z\bar{z}}=c_0+c_1 \log|z|+c_2\frac{(g_s M)^2}{[-i(\rho-\overline{\rho})]|z|^{4/3}},\label{eq:modKahler}}
where the last term dominates provided $\sigma z^{4/3} \ll 1$.
On the contrary, in the dilute flux regime ($\sigma z^{4/3} \gg 1$) the exponentially large $\sigma$ indicates  the extremely tiny gravitino mass.
For the supersymmetry breaking scale above TeV as implied by the LHC result to appear, the effective theory needs to be replaced by the Large Volume Scenario \cite{Balasubramanian:2005zx} (see also \cite{Crino:2020qwk}).
%From the scaling behavior of the warp factor $e^{-4A_0}$,  \cite{Blumenhagen:2019qcg} suggested that the modification above is equivalent to adding a new K\"ahler potential term 
%\dis{\Delta K_{\rm cs}= c_2 M^2 \frac{|z|^{2/3}}{[-i(\rho-\overline{\rho})][-(\tau-\overline{\tau})^2]}\label{eq:stwarp}}
%to  \eqref{eq:Kahler1}.
%While this new term contains $\tau$, it does not give a stringent effect on the behavior of $\tau$ since it as well as its derivatives with respect to $\tau$ are suppressed compared to those from \eqref{eq:Kahler1} for $g_s \ll 1$.
We also note that in the strongly warped regime, the Kaluza-Klein modes of the fields localized in the throat can be light by the redshift (for more discussion see \cite{Blumenhagen:2019qcg}).

 We finally note that far away from the throat, the warp factor is given by $e^{-4 A_0}\sim 4\pi {\alpha'}^2 g_s N/r^4$ such that the geometry is approximated by AdS$_5\times$S$_5$. 
 Here $r$ is the radial distance from the tip of the throat and $N$ is the total D-brane charges stored in the fluxes and the branes, thus $N\simeq MK$.
 This shows that the throat size in the string length unit is $(r_{\rm throat}/\ell_s)^4\sim g_s MK$ in the modified Einstein frame (or $\sim MK$ in the Einstein frame).

\section{Saddle point problem in dilaton stabilization}
\label{sec:EFT}

\subsection{Constraints on the dilaton-conifold modulus system}

In the GKP solution, the dilaton $\tau$ is stabilized at the  minimum satisfying $D_\tau W=0$ and $D_a W=0$.
 As the GVW superpotential is given in the form of  $W_{\rm GVW}=A(z, t^i)-\tau B(z, t^i)$, i.e., linearly dependent on $\tau$, and the K\"ahler potential is given by   \eqref{eq:Kahlertau}, the condition $D_\tau W_{\rm GVW}=0$ is written as
 \dis{D_\tau W_{\rm GVW}=-\frac{1}{\tau-\overline{\tau}} c \int_X \Omega \wedge \overline{G}_3=-\frac{1}{\tau-\overline{\tau}}(A-\overline{\tau}B)=0,}
 or $\tau=\overline{A}/\overline{B}$ in which the complex structure moduli take the  stabilized values.
 Meanwhile, by the condition $D_z W_{\rm GVW}=0$ the conifold modulus $z$ is stabilized at the exponentially small value 
  \dis{z \sim {\rm exp}\Big[-\frac{2\pi}{g_s M}\Big(K+f_1 K^0-\frac12\frac{f_1-f_1^*}{f_0-f_0^*}(K_0+2f_0 K^0)\Big)\Big].} 
 We note that the last term comes from $K_z W$ in the dilute flux limit, where the K\"ahler potential is given by \eqref{eq:Kahler1}.
 This term  vanishes for $|f_0-f_0^*| \sim {\cal O}(1)$ and $f_1-f_1^*=0$, as   supported by a concrete example considered in  \cite{Blumenhagen:2016bfp}.
 The well known value of $z$ corresponds to this case so hereafter we assume that $f_1$ is real such that $D_z W_{\rm GVW} = \partial_z W_{\rm GVW}+{\cal O}(z)$. 
 In the strongly warped regime, the dimensional reduction of the Type IIB supergravity action just replaces $G_{z\overline{z}}$ by \eqref{eq:modKahler}, so the minimum condition  $\partial_z W_{\rm GVW} \simeq 0$ is still valid \cite{Douglas:2007tu}. 
 For the string higher loops to be ignored in the effective theory, $g_s < 1$, or equivalently $-i(\tau-\overline{\tau})> 1$ is required.
 However, $g_s$ cannot be arbitrarily small by another effective theory validity condition that the size of $S^3$ at the tip of the deformed conifold is   larger than the string length scale \cite{Blumenhagen:2019qcg},%(see \eqref{eq:warp}) 
 \dis{R^2_{S^3} \sim e^{-2A(0)}|\epsilon|^{4/3} \sim \alpha' g_s |M| \gg \alpha',}
 or $g_s|M| \gg 1$.

 Now let us naively assume that the conifold modulus $z$ is the only complex structure modulus.
 While it is sensible only if the effects of other complex structure moduli are negligible, this illustrates the potential issues caused by  the linear dependence of $W_{\rm GVW}$ on $\tau$ and the parametric constraints discussed above. %, and the consistency with the KKLT scenario $W_0 \sim e^{-a \rho}$. 
 Suppose $K^0=K_0=0$ such that $W_{\rm GVW}$ does not contain a (constant)$\times \tau$ term, i.e., $B$ does not have a constant term,
  \dis{&A=c\Big(w_0+w_1 z+M\frac{z}{2\pi i}\log z\Big),\quad\quad B=c K z,\label{eq:Wtrial}}
%  \dis{W_{\rm GVW}=c\Big[w_0+w_1 z+M\frac{z}{2\pi i}\log z  -\tau z K\Big],}
  where $w_0=M_0+2f_0 M^0+f_1M$ and $w_1=f_1 M^0$.
  At the  minimum, conditions $D_z W_{\rm GVW}=0\simeq \partial_z W_{\rm GVW}$ and $D_\tau W=0$ give
  \dis{&\frac{M}{2\pi i}\log z=\tau K-w_1-\frac{M}{2\pi i},
  \\
  & z K\overline{\tau}=w_0+w_1 z+M\frac{z}{2\pi i}\log z = w_0 -\frac{M}{2\pi i}z+ z K\tau.}
  When Im$(\tau)$ is stabilized at $1/g_s$, they give the relation
  \dis{\frac{g_s}{2\pi}= \frac{2 K z}{2 \pi i w_0-M z}.\label{eq:gsexc}}
  We note that $w_0$ should be nonzero and indeed not be suppressed compared to $Mz$  since otherwise \eqref{eq:gsexc} gives $|z|\sim {\rm exp}[-(2\pi/g_s)(K/M)]\simeq e^{1/2}$, which contradicts to the exponentially small $|z|$. 
  Thus, we find that the absence of a (constant)$\times \tau$ term results in the exponentially small $g_s\sim |z|$.
  Such an exponentially small $g_s$ calls for the exponentially large $M$ and modestly large $K$ to satisfy $g_s |M| \gg 1$ and $(2\pi/g_s)(K/M)>1$ (for $|z|\ll 1$), respectively.
  In this case, a coefficient  $c=g_s^{3/2}(4\pi)^{-1/2}$ \cite{Conlon:2005ki} becomes exponentially small hence $W_{\rm GVW} \simeq c w_0$ can be as small as $e^{-a \sigma}(a\sigma)$ (or $\sigma\sim (a g_s)^{-1} K/M >1$) as required by the KKLT scenario. 
%  \footnote{One may worry about the additional tuning issue resulting from the exponentially large $M$ which appears in $w_0$.
%  For real but nonzero $f_1$, to make $g_s$ in \eqref{eq:gsexc} real the phases of $2\pi i w_0$ and $z$, which are determined by Re$(w_0)/{\rm Im}(w_0)=(M_0+2 {\rm Re}(f_0)M^0+f_1 M)/{\rm Im}(w_0)$ and $w_1=f_1 M^0$, respectively, must coincide.
% On the other hand, this is not a serious issue when $f_0$ is pure imaginary and $f_1=0$.
% Thus, the fine tuning issue concerning the phase is a model dependent one.}  
 Meanwhile, flux charges $M$ and $K$ are constrained by  the tadpole cancellation condition \eqref{eq:tadpole}, or $MK=-N_{D_3}+(1/2)N_{O_3}+\chi(X)/24$ then the exponentially large positive number $MK$ requires the exponentially large positive local charges in r.h.s.
 Since $\chi(X)$ ranges between $-240 \leq \chi(X)\leq 1 820 448$ \cite{Klemm:1996ts} and $\log(1820448/24)\simeq 11$, the value of $\log(M) \sim -\log(g_s) \sim (2\pi/g_s)(K/M)$ smaller than 10 may be allowed.
 However, as pointed out in \cite{Carta:2019rhx}, the throat size in the Einstein frame $(r_{\rm throat}/\ell_s)^4\sim MK$ is required to be smaller than the overall internal volume size $(V_X/\ell_s)^{4/6} \sim -i(\rho-\overline{\rho})=2\sigma$.
  Then the exponentially large $MK$ can be excluded by the modest value of $\sigma$.
  \footnote{ Even if $M$ is not exponentially enhanced, the parametric constraints generically call for large $M$, which gives rise to the similar issue.
  Away from the condition $g_s |M| \gg 1$, for example, if the supersymmetry is broken by the mechanism described in \cite{Kachru:2002gs}, the metastability of the uplifting term imposes (the number of $\overline{\rm D3}$ branes)/$|M|<0.08$.
  Then  given an integral number of $\overline{\rm D3}$ branes $M$ is required to be large.
  Since  $K$ is also not small as imposed by $(2\pi/g_s)(K/M)>1$, the K\"ahler modulus $\sigma$ needs to be  fixed at a large value to satisfy $MK<\sigma$.
  For example, we may consider (the number of $\overline{\rm D3}$ branes)=$2$, $g_s=0.1$, $M=30$, and $K=2$, which leads to $MK=60$.
  Then $e^{-a \sigma} <e^{-60 a}$ so for $a \sim {\cal O}(1)$, we have a very tiny ratio $m_{3/2}/m_{\rm Pl}\lesssim e^{-60}\simeq 10^{-26}$.
  In this way, the consistency conditions may require   the very large internal volume   and the extremely light gravitino mass.
   Then the low energy effective theory may be altered by the Large Volume Scenario \cite{Balasubramanian:2005zx}.
   }

Alternatively, we can introduce a (constant)$\times \tau$ term through nonzero $K_0,~K^0$ or the stabilization of other complex structure moduli $t^i$ \cite{Giddings:2001yu} to fix $\tau$   at the modestly large value.
Then  $g_s$ is perturbatively small but not exponentially suppressed.
 Especially, the existence of other complex structure moduli is appealing.
 While the requirement $W_{\rm GVW}\sim e^{-a \sigma}(a\sigma)$ is fulfilled  through the conspiracy of various stabilized field values, the low energy effective superpotential for $\tau$ after integrating out the complex structure moduli  is no longer linear in $\tau$.
 As pointed out in \cite{Choi:2004sx}, such a nontrivial $\tau$ dependence of the effective superpotential is crucial to prevent the destabilization of the vacuum.
 Even though the effective superpotential constructed by integrating out $z$ in the absence of other complex structure moduli is not linear in $\tau$ as well, it may not be enough to stabilize the potential.

\subsection{Saddle point problem}

 To see the instability issue discussed at the end of the previous section in detail, we first review  the saddle point condition  studied  in \cite{Choi:2004sx}, which observed the mass matrix of $\rho$ and $\tau$ with respect to the K\"ahler potential and the effective superpotential 
 \dis{&K=-3\log[-i(\rho-\overline{\rho})]-\log[-i(\tau-\overline{\tau})],
 \\
 &W_{\rm eff}=W_0(\tau)+C e^{i a \rho}.\label{eq:rhotaueff}}
  Here the superpotential term $W_0$ as a function of $\tau$ is constructed from $W_{\rm GVW}$ by integrating out $z$ as well as $t^i$.
 This assumes that the complex structure moduli including $z$ are heavier than $\tau$, and in fact, $m_z \gg m_\tau$ is generic in the dilute flux limit.
 This can be easily seen from the potential generated by the K\"ahler potential \eqref{eq:Kahler1} and the superpotential \eqref{eq:Wtrial}.
 From \eqref{eq:v''}, the $z$ mass is given by
 \dis{m_z^2\simeq e^K G^{z{\overline z}}\partial_z^2 W_{\rm GVW} \partial_{\overline z}^2 W_{\rm GVW} \simeq \frac{c^2  g_s M^2}{32 \pi \sigma^3 |z|^2\log|z|^{-2}},}
 so for $|z|\ll 1$, $z$ is very heavy.
  Including a (constant)$\times\tau$ term in $W_{\rm GVW}$ does not affect this result.
  In the strongly warped regime the modification of the K\"ahler metric \eqref{eq:modKahler} makes $z$ light for $|z|\ll 1$, but still heavier than $\rho$.
 Moreover, in the presence of the fields localized in the throat  the Kaluza-Klein modes  can be even lighter than $z$  so they also need to be taken into account in the effective theory for $(\rho, \tau)$  \cite{Blumenhagen:2019qcg}.
  Throughout this article, we assume that light fields localized in the throat do not exist or are irrelevant to the stabilization of $z$ and $\tau$.
  \footnote{In the example we consider, i.e., $z$ and $\tau$ are only relevant fields in GKP, the Kaluza-Klein modes of $\tau$ spoil the 4-dimensional effective theory when $m_\tau < m_z$ as $\tau$ behaves as if it were localized in the throat.
  This indeed gives more stringent condition $m_\tau > m\sigma\sim (a\sigma) m_{3/2}$ than our conclusion $m_{\tau} > m_{3/2}$.
   }
 
 To make our discussion simple, we consider the special form of $W_{\rm eff}$ where every term in $W_0$ has the same phase as $C$ and $a$ is real.
 By doing this, $\rho$ and $\tau$ have  purely imaginary values such that the pseudoscalars are fixed to zero at the supersymmetric extremum.
 From \eqref{eq:rhotaueff}, the second derivative matrix of the potential for the  scalars $\sigma={\rm Im}\rho$ and $s={\rm Im}\tau$ is given by  \cite{Choi:2004sx}
 \dis{V_s'' =\frac{|dW_{\rm 0}/ds|^2}{8 s \sigma^5}\left(
\begin{array}{cc}
3 s^2(7+10 a \sigma +4 (a \sigma)^2) & 3 s \sigma (3+2 a\sigma-2\gamma) \\
3 s \sigma (3+2 a\sigma-2\gamma) & \sigma^2(1-2\gamma+4 \gamma^2) \\
\end{array}\right),}
where 
\dis{\gamma=\frac{i s (d^2 W_0/d\tau^2)}{(d W_0/d\tau)}=\frac{ s (d^2 W_0/ds^2)}{(d W_0/ds)}.}
Note that $\gamma$ is real in our phase assignment.
The corresponding matrix for the pseudoscalars has the same structure \cite{Choi:2004sx}
 \dis{V_a'' =\frac{|dW_{\rm 0}/d s|^2}{8 s \sigma^5}\left(
\begin{array}{cc}
3 s^2(3+ 6 a \sigma +4 (a \sigma)^2) & 3 s \sigma (1+2 a\sigma-2\gamma) \\
3 s \sigma (1+2 a\sigma-2\gamma) & \sigma^2(1-2\gamma+4 \gamma^2) \\
\end{array}\right),}
which is identical to $V_s''$ for $a\sigma \gg 1$, so we focus on the scalar part $V''_s$ only.
Then the matrix $V''_s$ does not have a saddle point provided det$(V''_s)>0$, or
%\dis{|\gamma|>&\frac{1}{4(7+10 a \sigma+4(a\sigma)^2-3\cos^2\delta)}\Big[(-11-2 a\sigma+4(a\sigma)^2)\cos \delta\\
%&+\sqrt{(7+10 a \sigma+4(a\sigma)^2)[8(10+13 a \sigma+4(a\sigma)^2)-(17+14 a\sigma-4 (a\sigma)^2)\cos^2\delta]}\Big],}
\dis{\gamma>\frac{5+ 4 a\sigma}{4(2+a\sigma)},\quad\quad \gamma<-\frac{2+a\sigma}{1+2a\sigma}.}
%In the limit of $a\sigma \to \infty$, $|\gamma|$ has a lower bound  $(\cos\delta+\sqrt{\cos^2\delta +8})/4$.
Thus, we conclude that for the potential to be stabilized,   $|\gamma|> {\cal O}(1)$ should be imposed \cite{Choi:2004sx}.
If $W_0$ is linear in $\tau$, i.e., $W_0=A-\tau B$ with $A,~B$   constants, $\gamma=0$ so the potential has an unstable direction at the supersymmetric field values.

Now   we go back to the case of the superpotential given in the form of $W_{\rm GVW}=A(z)-\tau B(z)$, where
\dis{A&=c\Big(w_0+w_1 z+ M\frac{z}{2\pi i}\log z\Big),\quad\quad
B=c\Big((K+ f_1 K^0)z + w_2 \Big).
\label{eq:GVW1}}
To make our discussion simple, we restrict our discussion to the case where $w_0$ is pure imaginary, $w_1$ is pure imaginary or zero, and $w_2$ is real.
By doing this, $z$ is stabilized at the real value while $\tau$ is fixed at $i/g_s$, i.e., pure imaginary value.
 When the effects of complex structure moduli other than $z$ are completely negligible, we have  $w_0=M_0+ 2f_0 M^0+f_1M$, $w_1=f_1 M^0$ and $w_2=K_0+2f_0K^0$ so  by taking $M_0=K^0=0$ and $f_1=0$ we have real $(w_0, w_2)$ and vanishing $w_1$. 
However, this is not consistent with the  exponentially small $c w_0$ (hence $W_0$) for the following reason.
From the minimum conditions $D_z W_{\rm GVW}=D_\tau W_{\rm GVW}=0$ one finds that
\dis{\frac{g_s}{2\pi}= \frac{2 K z+w_2}{2 \pi i w_0-M z},}
in which $g_s$ is, in general, not exponentially tiny   but estimated to be $g_s \sim w_2/(i w_0)$ as $z$ is exponentially suppressed. 
Then after stabilization of $z$ and $\tau$, we have $W_0\simeq c(w_0-ig_s^{-1} w_2)=2 c w_0$, the absolute value of which is required to be comparable to $(a\sigma)e^{-a \sigma}$ and at the same time, $w_2 \sim (i w_0)g_s < iw_0$ is satisfied.
This cannot be achieved with the integer value of $w_2=K_0$.
 For this reason, we will not attach to the explicit forms of coefficients $(w_0, w_1, w_2)$ in order to accommodate the superpotential as a function of $(z, \tau)$ induced by integrating out $t^i$ but still linear in $\tau$, i.e., coefficients are determined by other complex structure moduli effects as well as the flux quanta.
\footnote{One may include the non-perturbative term $Ce^{ia \rho}$ in $c w_0$.
Then  $\tau$ is stabilized at the same value as that obtained from $W_{\rm eff}$ after the $\rho$ stabilization.
This does not change the stabilized value of $g_s$ much provided $c w_0 \gtrsim Ce^{-a\sigma}$.}
Our discussion on the stability shows that even  $(w_0, w_1, w_2)$ are given by free parameters and we allow the tuning of parameters to satisfy $z<w_0<1$, the stability issue rules out such possibility.
That means integrating out complex structure moduli must provide a more generic function of $\tau$, which will be consistent with the flux condition discussed in section \ref{sec:Condition}.

To see the stability condition, we consider the condition $D_z W =0 \simeq \partial_z W_{\rm GVW}$,
\dis{w_1 +\frac{1}{2\pi i}(1+\log z)M-\tau (f_1 K^0+K)=0,}
 from which $z$ is expressed as a function of $\tau$, 
 \dis{z={\rm exp}\Big[\frac{2\pi i}{M}\Big(\tau(f_1 K^0+K)-w_1\Big)-1\Big].\label{eq:z}}
 Putting this back to $W_{\rm GVW}$  the  effective superpotential in the absence of the non-perturbative effect is written as 
\dis{W_0 (\tau) \simeq c\Big(w_0-\tau w_2 -\frac{M}{2\pi i}{\rm exp}\Big[\frac{2\pi i}{M}\big(\tau(f_1 K^0+K)-w_1 \big)-1\Big]\Big).\label{eq:Weff}}
 It is remarkable that $W_0(\tau)$ here has an exponential term, which looks like the non-perturbative effect under $s={\rm Im}(\tau)=1/g_s$.
 As pointed out in \cite{Blumenhagen:2016bfp}, this shows that the shift symmetry $\tau \to \tau+ \theta$ of the K\"ahler potential is broken to the discrete one through the coupling between $\tau$ and $z$ induced by the $H_3$ flux (the term $-\tau (f_1 K^0+K)z$ in \eqref{eq:GVW1}).
 More explicitly, if $w_2=0$ ($K_0=K^0=0$), the discrete symmetry $\tau \to \tau+(M/K)n$ with $n \in \mathbb{Z}$ remains as a remnant of the shift symmetry.
 This indeed is originated from the monodromy property of the conifold modulus $z$, ${\cal F}_z \to {\cal F}_z+1$ while $z \to e^{2\pi i} z$ (see \eqref{eq:zcycle}), which is a part of Sp$(2(h^{2,1}+1))$ symplectic structure of the holomorphic 3-form $\Omega$  as given by \eqref{eq:Omega} \cite{Candelas:1990pi}.
 In \eqref{eq:GVW1}, we immediately find that when $K_0=K^0=0$,   the shift $\log z \to \log z+2\pi i$ from the monodromy $z \to e^{2\pi i}z$ is compensated by the discrete shift $\tau \to \tau+(M/K)$.

After the stabilization of $\tau$ ($s=g_s^{-1}$), we obtain  
\dis{\gamma=-\frac{2\pi}{g_s M}\frac{(f_1 K^0+K)^2 z}{w_2+(f_1 K^0+K)z},}
where $z$ is given by \eqref{eq:z}, which is an exponentially suppressed value.
Thus, when $w_2$, or a (constant)$\times \tau$ term in $W_{\rm GVW}$ is non-vanishing and even not suppressed compared to $(f_1K^0+K)z$,  $|\gamma|$ is exponentially small. 
Then  the potential has a saddle point by the mass mixing   with $\rho$  so becomes unstable.  
For $w_2=0$, $\gamma=-(2\pi K/g_s M)$ is noting more than the exponent of $|z|$ then $|\gamma|>1$ is satisfied.
As we have seen in the discussion below \eqref{eq:gsexc}, however, in this case $g_s$ becomes exponentially small, which can be excluded by the condition $MK < \sigma$.

\section{Interpretation of the stability condition}
\label{sec:Condition}

\subsection{Comparison with gravitino mass}

The physical interpretation of the stability condition $|\gamma|>1$ is clear by comparing the dilaton mass with the gravitino mass $m_{3/2}$.
 Since the Lagrangian is written as
 \dis{-G_{\tau{\overline \tau}}\partial_\mu \tau \partial^\mu \overline{\tau}-V(\tau)+\cdots= -\frac{1}{4 s^2} (\partial s)^2-\frac12 \frac{\partial^2 V}{\partial s^2}\Big|_{\rm min} s^2+\cdots,}
 the $(22)$ component of $V''_s$ is related to the $(22)$ component of the mass matrix by
 \dis{M^2_{ss}=2 s^2 \frac{\partial^2 V}{\partial s^2}\Big|_{\rm min}=\frac{s}{4 \sigma^3}\Big(\frac{d W_0}{d s}\Big)^2(1-2\gamma+4\gamma^2). \label{eq:Mss}}
 Meanwhile, the supersymmetric minimum conditions $D_\tau W_{\rm eff}=0=D_\rho W_{\rm eff}$ read
\dis{Ce^{-a\sigma}=-\frac{3}{3+2 a \sigma}W_0,\quad\quad
s\frac{dW_0}{ds}=\frac{a\sigma}{3+2 a\sigma} W_0,}
from which the gravitino mass is given by
\dis{m_{3/2}^2=e^K|W|^2=\frac{1}{16 \sigma^3 s}|W_0+Ce^{- a \sigma}|^2=\frac{s}{4\sigma^3}\Big(\frac{d W_0}{d s}\Big)^2.\label{eq:m3/2}}
Thus, in the limit   $\gamma \to 0$,  $M^2_{ss}$ becomes $m_{3/2}^2$.
 When the stability condition $|\gamma|>1$ is satisfied, $M^2_{ss}$ gets considerably larger than $m_{3/2}^2$.
 Since $m_\sigma \sim (a\sigma) m_{3/2}> m_{3/2}$, this indicates that unless $|\gamma| \gg 1$, $\tau$ cannot be decoupled from $\rho$ and the mass mixing caused by the non-perturbative effect should not be neglected.
The original KKLT scenario assumed $|\gamma|\gg 1$ for the decoupling of $\tau$ from $\rho$.
 In order to see how the non-perturbative effect changes $M_{ss}^2$, let us for a moment turn off the non-perturbative effect by taking  $a \to \infty$, and observe the property of $M_{ss}^2$  generated by $W_0$ only.
 In this case, the potential has a no-scale structure $V=e^K G^{\tau\overline{\tau}}D_\tau W D_{\overline{\tau}}\overline{W}$ and we have a simple relation $2s(dW_0/ds)=W_0$ at the  minimum.
 From this we obtain 
 \dis{\frac{d^2V}{d s^2}\Big|_{\rm min}=\frac{|dW_0/ds|^2}{8\sigma^3 s}|1+2\gamma|^2,}
   which is positive definite for any value of $\gamma$ while this is not the case in the presence of the non-perturbative effect, as can be seen from the $(22)$ component of $V''_s$.
   
 Since $W_0$ is obtained by integrating out the complex structure moduli, the properties of $W_0$ including $\gamma$ encode how the complex structure moduli are stabilized  by $W_{\rm GVW}$.
 From $W_{\rm GVW}=A(t^a)-\tau B(t^a)$ where $t^a=(z, t^i)$, the  minimum conditions read
 \dis{&D_a W_{\rm GVW}=(\partial_a A-\tau \partial_a B)+ \partial_a K(A-\tau B)=0,
 \\
 &D_\tau W_{\rm GVW}=-\frac{1}{\tau-\overline{\tau}}(A- \overline{\tau} B)=0,}
 or equivalently,
 \dis{A=\overline{\tau} B,\quad\quad \partial_a A-\tau \partial_a B= -\partial_a K (\tau-\overline{\tau})B.}
Then using \eqref{eq:v''} we obtain
 \dis{m_{\tau}^2 &= G^{\tau {\overline \tau}}\partial_\tau \partial_{\overline\tau} V
 \\
 &=-(\tau-\overline{\tau})^2 e^K \Big(G^{a{\overline b}}(\partial_a B+\partial_a K B)(\partial_{\overline b} \overline{B}+\partial_{\overline b} K {\overline B})+|B|^2\Big).\label{Eq:taumass}} 
 This can be compared to
 \footnote{While $m_{3/2}$ here corresponds to the  $a\to \infty$ limit of \eqref{eq:m3/2} and its numerical value is not much changed under this limit as $W_0 \sim e^{-a\sigma}(a\sigma) \gg C e^{-a\sigma}$, the supersymmetry breaking occurs in a different way. 
 In the GKP solution which employs $W_{\rm GVW}$, the potential has a no-scale structure.
 Then it is stabilized at the Minkowski vacuum through the cancellation between  $G^{\rho {\overline \rho}}|D_\rho W_{\rm GVW}|^2$ and $3|W_{\rm GVW}|^2$, which implies that the supersymmetry is broken by nonzero $D_\rho W_{\rm GVW}$ (for the possible problem in nonzero $W_0$, see, \cite{Sethi:2017phn}).
 On the other hand, in the KKLT scenario all fields are stabilized at the supersymmetric AdS minimum satisfying $D W=0$ and the  supersymmetry is broken by the uplift.
 Since the original KKLT assumed $\tau$ as well as $t^a$ are heavy enough to be decoupled from the KKLT scenario, their stabilized values are not much affected by the small non-perturbative effect.
% This justifies our discussion on the decoupling condition in terms of $W_{\rm GVW}$ where the non-perturbative effect is ignored.
  }
 \dis{m_{3/2}^2=e^K|A-\tau B|^2=-(\tau-\overline{\tau})^2 e^K|B|^2,}
 which is exactly the second term of \eqref{Eq:taumass}.
 That is, in the absence of the coupling to the complex structure moduli  $m_\tau^2$ which is equivalent to $M_{ss}^2$ is given by $m_{3/2}^2$.
 This can be found in \eqref{eq:Mss} as well : $\gamma$ is generated by the interaction between $\tau$ and the complex structure moduli so when the interaction is turned off, $\gamma$ becomes zero, and we obtain $M_{ss}^2=m_{3/2}^2$. 
 Moreover, the first term of \eqref{Eq:taumass} is positive definite, which is consistent with the positive definiteness of $d^2V/ds^2$ obtained by turning off the non-perturbative effect.

 In terms of the discussion above, the problem in considering the conifold modulus $z$ as the only complex structure modulus can be understood in the following way.
 Since
 \dis{&B=c\Big((K+ f_1 K^0)z + w_2 \Big),\quad\quad
\partial_z B=c(f_1 K^0+K),}
and the K\"ahler potential is given by \eqref{eq:Kahler1} with $t^i$  turned off, the first term of \eqref{Eq:taumass} becomes
\dis{G^{z{\overline z}}|\partial_z B+\partial_z K B|^2 \simeq \frac{2\pi i (f_0-f_0^*)}{\log|z|}|\partial_z B|^2 \simeq \frac{2\pi i (f_0-f_0^*)}{\log|z|}c^2|f_1K^0+K|^2.  \label{eq:CSMcont}}
Since  $-\log|z|=(2\pi/g_s)(f_1K^0+K)/M > 1$ and $B\simeq cw_2$ unless $w_2 < (f_1 K^0+K)z$, this term is typically suppressed by $-\log|z|$ compared to $m_{3/2}^2$.
It corresponds to the case $|\gamma|<1$ in which the potential is destabilized when the non-perturbative effect and the mass mixing are taken into account additionally. 
If  $w_2 < (f_1 K^0+K)z$, e.g., $K_0=K^0=0$, $B\simeq (f_1 K^0+K)z$ hence \eqref{eq:CSMcont} becomes larger than $m_{3/2}^2$.
While this allows $|\gamma|>1$ and the potential is stable against the mass mixing, as we have seen, this   requires the exponentially small $g_s$ and exponentially large $M$, easily violating the condition $MK <\sigma$.
In the highly warped regime, the K\"ahler metric is replaced by \eqref{eq:modKahler}, to give
 \dis{G^{z{\overline z}}|\partial_z B+\partial_z K B|^2&\simeq
c^2\Big(c_0+c_1 \log|z|+c_2\frac{(g_s M)^2}{2\sigma|z|^{4/3}}\Big)^{-1}| f_1 K^0+K |^2
\\
&\simeq \frac{2c^2 |z|^{4/3} \sigma}{c_2 (g_s M)^2} | f_1 K^0+K|^2,}
which is typically smaller than $|B|^2$ unless $w_2 < (f_1 K^0+K)z$ thus the potential is easily destabilized by the mass mixing.

\subsection{Flux condition} 

 We have seen that $\tau$ is  stabilized without the saddle point provided $\tau$ is considerably heavier than $m_{3/2}$, or equivalently
 \dis{G^{a{\overline b}}(\partial_a B+\partial_a K B)(\partial_{\overline b} \overline{B}+\partial_{\overline b} K {\overline B})>|B|^2. \label{eq:Bcond}}
 Since $\tau$ as well as $t^a=(t^i,~z)$ are stabilized by the fluxes,  such stabilization condition  can be translated into the condition on the fluxes.
 To see this,  we first note that $B$ in $W_{\rm GVW}$ corresponds to 
 \dis{B=c \int_X \Omega \wedge H_3.}
 Since the complex structure moduli $t^a$ are contained in $\Omega$ and $\partial_a \Omega = k_a \Omega +\chi_a$, i.e. a combination of the $(3,0)$- and the primitive $(2,1)$ forms, we have
 \footnote{For the strongly warped regime, a factor $e^{-4A_0}$ is multiplied to each integrand, giving, e.g., \eqref{eq:modKahler} \cite{DeWolfe:2002nn}.}
 \dis{&\partial_a K= \frac{\int_X  \partial_a \Omega \wedge \overline{\Omega}}{\int_X \Omega\wedge \overline{\Omega}}=-k_a,
\quad\quad
G_{a\overline{b}}=-\frac{\int_X  \chi_a \wedge \overline{\chi}_{\overline b}}{\int_X \Omega\wedge \overline{\Omega}},\label{eq:kG}} 
 from which we obtain
 \dis{\partial_a B+K_a B=c \int_X \chi_a \wedge H_3.\label{eq:DaB}}
 We note that for
 \dis{D_a W=c \int_X \chi_a \wedge G_3 =0}
to be satisfied at the  minimum,  $G_3$ is required to be   imaginary self dual, so $(3,0)$ and $(1,2)$ components do not exist \cite{Giddings:2001yu}.
This however does not mean that $H_3$ does not have a $(1,2)$ component, but constrained by the condition $G_{(1,2)}=F_{(1,2)}-\tau H_{(1,2)}=0$ at the  minimum.
Otherwise, \eqref{eq:DaB} becomes zero such that \eqref{eq:Bcond} is violated, which means that the potential is destabilized. 
 Then $ H_{(1,2)}$ and $ H_{(0,3)}$ are constrained by \eqref{eq:Bcond}, which reads
 \dis{\frac{\int_X \chi_a \wedge H_{(1,2)} \int_X \chi_{\overline b}\wedge \overline{H}_{(2,1)}}{\int_X \chi_a \wedge \chi_{\overline b} } >  
 \frac{\Big|\int_X \Omega \wedge H_{(0,3)}\Big|^2 }{\int_X \Omega \wedge \overline{\Omega} }\label{eq:fluxcond}}   
 with the help of \eqref{eq:kG} and \eqref{eq:DaB}.
 One obvious way to satisfy this condition is demanding $|H_{(1,2)}| > | H_{(0,3)}|$.
Thus, the decoupling of the dilaton  from the low energy physics of the typical mass scale $m_{3/2}$ or $m_\sigma \sim (a \sigma)m_{3/2}$ is achieved when the $(1,2)$ component of $H_3$ dominates over the $(0,3)$ component.
 Since $W_{\rm GVW}=c\int \Omega \wedge G_3 = c\int \Omega \wedge G_{(0,3)}$, the low energy supersymmetry breaking is realized provided the $G_{(0,3)}$ component is small, say, $G_{(0,3)} \ll G_{(2,1)}$ \cite{Choi:2005ge}.
 The smallness of $H_{(0,3)}$ is consistent with such requirement while  there still remains the tuning to make $W_{\rm GVW}=A-\tau B = -(2i/g_s)B$  as small as $e^{-a \sigma}(a\sigma)$.
In \cite{Choi:2005ge}, it was also pointed out that the flux vacua satisfying $G_{(0,3)}/ G_{(2,1)} <\epsilon$ take up about $\epsilon^2$ of the total number of vacua \cite{Ashok:2003gk, Denef:2004ze, Giryavets:2004zr, Gorlich:2004qm}.
We expect that \eqref{eq:fluxcond}, or $|H_{(1,2)}| > | H_{(0,3)}|$ provides a more stringent constraint on the search for the flux vacua consistent with our universe.

 \section{Conclusions}
\label{sec:conclusion}

We have discussed the condition for the dilaton stabilization consistent with  the KKLT scenario.
The dilaton potential is generated by the flux-induced GVW superpotential, as it breaks the shift symmetry of the K\"ahler potential for the dilaton $\tau$.
Moreover, the fact that the GVW superpotential is linearly dependent on  $\tau$ may give rise to the saddle point problem.
That is, if $\tau$ does not couple to any complex structure moduli, the $\tau$ mass is simply given by the gravitino mass $m_{3/2}$.
Then when we take the non-perturbative effect of the K\"ahler modulus $\rho$  into account the supersymmetric solution is on the saddle point, hence the potential becomes unstable through the mass mixing between $\tau$ and $\rho$.
 While $\tau$ couples to the conifold modulus $z$, another essential ingredient of both the GKP solution and the KKLT scenario, this does not help to resolve the problem in the controllable parameter space.
 Alternatively, we can introduce the complex structure moduli $t^i$ other than $z$ in which the exponentially small $W_0 \sim e^{-a\sigma}(a\sigma)$ as required by the KKLT scenario is a result of the conspiracy of stabilized field values or fluxes.
 
 In summary, the saddle point problem is circumvented when $\tau$ is considerably heavier than $m_{3/2}$ through the coupling to the complex structure moduli $t^i$.
%The mechanism to achieve this can be understood in two ways.
In the low energy effective theory point of view,  the  effective superpotential $W_0(\tau)$ generated by integrating out $t^i$ as well as $z$ does no longer linearly depend on $\tau$, then an appropriate functional form of $W_0(\tau)$ makes $m_\tau \gg m_{3/2}$.
In terms of the GVW superpotential which  still linearly depends on $\tau$, this can be understood as a result of the complex structure moduli stabilization under the constraint $|H_{(1,2)}| > | H_{(0,3)}|$ on the flux.
  This is consistent with the flux condition $G_{(0,3)}\ll G_{(2,1)}$ for the low energy supersymmetry breaking, but more restrictive.
% Indeed, many discussions on the GKP solution and the KKLT scenario have focused on the conifold and assumed that other parts of the internal manifold are decoupled.
 It is remarkable that since $\tau$ linearly couples to any complex structure moduli, its stabilization provides the restriction from the unknown part of the internal manifold.
 Our study which connects the vacuum stability to the flux condition can be an example of such restriction.

 We also point out that as a by-product, our study can shed light on the cosmology of the multi-field axion inflation \cite{Kim:2004rp, Dimopoulos:2005ac}.
 The effective superpotential after integrating out $z$ contains the non-perturbative effect-like term, so after the stabilization the potential for  $C_0$, the pseudoscalar component of $\tau$ is given by the sum of the axion-like term and the polynomial term as well as their mixing. 
 The fixed scalar field value ${\rm Im}(\tau)=g_s^{-1}$ provides the axion decay constant.
 This implies that the features of the $(\rho, \tau)$ system also appear in the two-field axion inflation model with the perturbative correction \cite{Blumenhagen:2016bfp}.
 In this regard, we expect that our discussion on the stability provides some conditions on the multi-field axion inflation model, which also can be strongly constrained by the weak gravity conjecture \cite{ArkaniHamed:2006dz} (see also e.g., \cite{Cheung:2014vva, Brown:2015iha}).
  Moreover, the flux condition as well as the parametric restrictions on the internal manifold consistent with the dilaton and the moduli stabilization may be used to revisit  the inflation models using the throat, such as the DBI inflation model \cite{Silverstein:2003hf} (see also, e.g., \cite{Kecskemeti:2006cg, McAllister:2007bg, Chen:2008hz, Seo:2018abc, Mizuno:2019pcm}).

\subsection*{Acknowledgements}

MS is grateful to Kang-Sin Choi and Bumseok Kyae for discussion and comments    while this work was under progress.
%

%

%\newpage

\appendix

\renewcommand{\theequation}{\Alph{section}.\arabic{equation}}

%\section{Uncertainty for the infrared modes}
%\label{app:IRuncert}
%\setcounter{equation}{0}

\end{document}